\documentclass[aps,prd,preprint,groupedaddress,showpacs,11 pt]{revtex4-1}
\usepackage{amsmath,amstext,amsbsy,amssymb}
\usepackage{bm}
\usepackage{color}

\newcommand{\ssA}{{\scriptscriptstyle{ A}}}
\newcommand{\ssV}{{\scriptscriptstyle{ V}}}

\newcommand{\ssR}{{\scriptscriptstyle{ R}}}
\newcommand{\ssH}{{\scriptscriptstyle{ H}}}

\newcommand{\ssL}{{\scriptscriptstyle{ L}}}

\newcommand{\ssM}{{\scriptscriptstyle{ M}}}

\newcommand{\ssW}{{\scriptscriptstyle{ W}}}

\newcommand{\ssl}{{\scriptscriptstyle{ \lambda}}}
\newcommand{\Li}{{\rm Li}}

\long\def\symbolfootnote[#1]#2{\begingroup%
\def\thefootnote{\fnsymbol{footnote}}\footnote[#1]{#2}\endgroup}

\begin{document}

\title{Nonlinear  Chiral Plasma Transport  in Rotating Coordinates}

\author{\"{O}mer F. Dayi}
\email{dayi@itu.edu.tr }
\author{Eda Kilin\c{c}arslan}
\email{kilincarslan@itu.edu.tr }
\affiliation{%
	Physics Engineering Department, Faculty of Science and
	Letters, Istanbul Technical University,
	TR-34469, Maslak--Istanbul, Turkey}


\begin{abstract}
The nonlinear transport features of inhomogeneous chiral plasma in the presence of electromagnetic fields, in rotating coordinates are studied
within the relaxation time approach. The chiral  distribution functions up to  second order in the electric field in rotating coordinates and the derivatives of chemical potentials are established by solving the Boltzmann transport equation. First, the vector and axial current densities in the weakly ionized chiral plasma for vanishing magnetic field are calculated.  They involve the rotational analogues  of  the Hall effect as well as several new terms arising from the Coriolis and fictitious  centrifugal forces. Then in the short relaxation time  regime the angular velocity and electromagnetic fields are  treated as perturbations. The current densities are obtained  by retaining the terms up to  second order in perturbations.  The time evolution  equations of the inhomogeneous chemical potentials  are derived by demanding that collisions conserve the particle number densities. 
	
\end{abstract}

\maketitle

\newpage

\section{Introduction}
\label{int}

The classical formulation of physical systems may possess some symmetries which do not survive in the quantum regime. One of the prominent
examples has been the chiral symmetry of the massless Dirac equation which is broken  in the presence of  electromagnetic fields. It is known as the  chiral anomaly which promotes
 some anomalous transport  phenomena of chiral particles like 
the chiral magnetic effect \cite{kmw,fkw,kz} and the chiral separation effect \cite{mz,jkr}. 
There are  similar anomalous effects when the chiral particles are subject to rotations: The 
chiral vortical effect \cite{ss} and the local (spin) polarization effect \cite{lw,bpr,glpww}. 
These phenomena can be viewed  as noninertial effects resulting from the observation of  particles in a rotating coordinate frame.

The chiral transport phenomena manifest itself  in diverse physical systems. We focus on the  relativistic chiral plasma  transport. In the early  Universe electroweak plasma can occur due to the excess of  right-handed electrons over left-handed ones \cite{js}. Then,   primordial magnetic field can be generated due to the chiral vorticity and chiral magnetic effects \cite{tvv}. This was also  studied   within the kinetic theory by dealing with  the distribution functions which are computed perturbatively in  the Fourier transformed electromagnetic fields  \cite{bp}. The chiral plasma can also be created in relativistic heavy-ion collisions where  some experimental evidences of the chiral anomalous effects may exist \cite{kmw,fkw}. 

When one studies  the dynamics of a plasma, it is not possible to consider the dynamics of  individual particles, instead one should deal with the distribution functions. An intuitive formulation of the chiral plasma is offered by the semiclassical kinetic theory after being able to 
 incorporate the chiral anomaly into  it \cite{soy,sy}. Kinetic theory relies on the Boltzmann transport equation which is known to be suitable in examining the properties of weakly ionized plasma \cite{KrTr}. In fact we deal with a hot chiral plasma which is roughly  neutral.
 Hence  particles and antiparticles contribute on an equal footing to the  particle number and current densities.

The semiclassical kinetic theories of the Dirac and Weyl particles in the presence of   electromagnetic fields,
in a rigidly rotating coordinate frame   were established  within the differential forms method in Ref.\cite{dky}, by generalizing the  symplectic two-form of the nonrelativistic particles in rotating coordinates to embrace the relativistic particles  by retaining rotations  nonrelativistic. There exists another approach
 \cite{cpww} where the chiral transport was addressed starting from a relativistic kinetic equation defined in terms of the Wigner functions \cite{qBe}. By integrating out the zeroth component of 4-momentum, they obtained a three dimensional kinetic equation. These two formalisms differ mainly how they treat the resemblance between the Aharonov-Bohm phase caused by the  magnetic field,  $\bm B,$  and the Sagnac effect due the angular velocity, $\bm \Omega;$ namely the symmetry between  $\bm B$ and  $2{\cal{E}}\bm \Omega ,$  where ${\cal{E}} $ denotes the relativistic energy. In the  approach of \cite{dky} this symmetry is respected, however in the formalism of \cite{cpww} it is violated in some of the equations of motions and the measure of the phase space.
 
 We employ the formalism of Ref.\cite{dky}, where
the Boltzmann equation    without collisions has been  taken into account in calculating the chiral currents.  
We would like to explore  the nonlinear transport properties of the chiral plasma in rotating coordinates  within the relaxation time approach.
The effects arising in a frame  which rotates with respect to the laboratory frame correspond to the effects caused by the  vorticity in fluids.   Now, the chemical potentials are inhomogeneous and  may  in general be time dependent. The equations governing their evolution in time will be delivered by demanding that the collisions conserve  particle number densities.    

We focus on the nonlinear chiral transport of inhomogeneous media generated by rotations in the presence of electric and magnetic fields.   We calculated the distribution functions by  solving the Boltzmann equation of the chiral particles in rotating coordinates  up to second order in  electric field and the linear velocity arising from rotations. By employing  these distribution  functions the vector and axial current densities as well as  the equations governing the time evolution of chemical potentials are computed. We will see that besides the  Ohm's law,  the analogue of the Hall effect for rotation and currents related to the fictitious centrifugal force, there are several new terms in the chiral current densities.  A similar approach was presented in Ref.\cite{gorbaretal}
for the neutral plasma  in the absence of rotations.

In the next section we first  briefly review the semiclassical kinetic theory of chiral particles in a rigidly rotating frame. Then the relaxation time method which we adopt to introduce collisions is presented. Section \ref{SSBE} is devoted to the calculation of distribution functions up to second order in the derivatives of chiral chemical potentials, the electric field and the linear velocity due to the rotation of the coordinate frame. The definitions of particle number and current densities are given in Sec. \ref{pnc}. The distribution functions are employed to establish the current densities arising
from the rotations and  electric field in the absence of  magnetic field in Sec. \ref{b0}. We then treated  the electromagnetic fields, angular velocity and the derivatives of chemical potentials as perturbations in the short relaxation time regime. The distribution functions acquired within this weak fields approximation are employed to  calculate the vector and axial current densities as well as the time evolution equations of chemical potentials in Sec. \ref{wfa}. We then summarized and discussed the results in the last section.

\section{Review of the semiclassical approach}
\label{rev}

We would like to recall briefly the semiclassical kinetic theory of the Weyl particles under the influence of electromagnetic fields, in rotating coordinates, following the formalism of Ref.\cite{dky}.
We set the speed of light $c=1$ and the Boltzmann constant $k=1.$ We deal with the chiral particles and antiparticles in
coordinates rotating with the  constant angular velocity  $\bm \Omega ,$ in the presence of the  electromagnetic fields $\bm E ,\ \bm B .$ 
Although the nonrelativistic rotations  obeying the condition $|\bm \Omega \times \bm x|\ll c$ are considered, the centrifugal effects are not ignored.
The Maxwell equations in rotating coordinates  are given as 
\begin{eqnarray}
\bm \nabla \cdot \bm E^\prime & = &4\pi q n,\nonumber\\
\bm \nabla \times \bm E^\prime & = &- \frac{\partial \bm B}{\partial t},\nonumber\\
\bm \nabla \cdot \bm B & = & 0, \nonumber \\
\bm \nabla \times \bm B & = &4\pi q \bm j + \frac{\partial \bm E^\prime}{\partial t}, \label{Max}
\end{eqnarray}
where the electric field in the rotating frame is 
\begin{equation}
\label{rote}
\bm E^\prime =  \bm E + (\bm\Omega\times \bm x)\times \bm B .
\end{equation}
The electric and magnetic fields, $\bm E,\ \bm B,$ will evolve in plasma according to the Maxwell equations with
the particle number  and current densities $n,$  $\bm j,$ which must be acquired consistently. In fact the main objective of this work is to establish number and current densities according to kinetic theory  of the chiral particles and antiparticles which are labeled by $i:$  $q_p= q$ for a particle and  $q_a=-q$ for an antiparticle.  The
helicity states are denoted by   $\lambda =+1,-1,$ corresponding, respectively, to the right- and left-handed particles.
The Hamiltonian formalism  is provided with 
the extended symplectic two-form $(a,b,c=1,2,3)$
\begin{eqnarray}
(\omega_{t})_i^\ssl & = & {dp}_a \wedge {dx}_a {+ \epsilon_{abc}   \Omega_c x_b \nu_{ m}^\lambda {dx}_a \wedge {dp}_m }+ \frac{1}{2} \epsilon_{abc} (q_i B_b + 2{\cal E}_{i}^\lambda \Omega_b) {dx}_a \wedge {dx}_c \nonumber \\ 
&&+ \frac{1}{2} \epsilon_{abc} b^\lambda_{ic} {dp}_a \wedge {dp}_b 
-\nu^\lambda_{0ia} ( 1{- \frac{1}{2}  (\bm \Omega \times \bm x)^2}) {dp}_a \wedge dt \nonumber \\ 
&&+[q_i\bm E+ (\bm\Omega\times \bm x)\times (q\bm B +{\cal E}_{i}^\lambda\bm\Omega)]_a{dx}_a \wedge dt ,
\label{masslesstwoform} \nonumber
\end{eqnarray}
in terms of  the  Berry curvature    
$$
\bm{b}_i^\lambda =\lambda\mathrm{sign}(q_i) \hbar \frac{\bm p}{2p^3},
$$
and the ``canonical" velocity
\begin{eqnarray}
\label{nu0}
{\bm \nu}_{0i}^\lambda&=&\frac{\bm p}{p} + \lambda \mathrm{sign}(q_i)  \hbar \frac{\bm p}{2 p^3} ( \bm{\Omega}\cdot\bm{p} ) -\lambda \mathrm{sign}(q_i) \hbar \frac{\bm \Omega}{2 p} + \lambda \hbar q\frac{\bm p}{p^4} (\bm{B} \cdot \bm{p}) - \lambda\hbar q \frac{\bm B}{2 p^2}.
\end{eqnarray}
It is derived from  the (dispersion relation) Hamiltonian 
\begin{equation}
\label{disp}
{\cal E}_i^\lambda = p  -  p^2 (\bm{b}_i^\lambda \cdot \bm{\Omega}) - q_ip (\bm{b}_i^\lambda \cdot \bm B).
\end{equation}
In the semiclassical approach we keep the terms which are at most linear in the Planck constant. Obviously, these are up to the   Planck constants  appearing in the definition of momentum space volume: $(2\pi \hbar)^{-3}.$

The extended symplectic two-form $\omega_{t} $ was utilized to establish the  phase space measure (Pfaffian) and the time derivatives of phase space variables as follows,
\begin{eqnarray}
(\sqrt{\omega})_i^\lambda &=&1+  \ \bm{b}_i^\lambda \cdot (q_i\bm{B} + 2p \bm{\Omega} )
{ - \bm \nu_{0i}^\lambda \cdot (\bm x \times \bm \Omega) -  \hat{\bm p}  \cdot \bm b_i^\lambda(q_i \bm B \cdot (\bm x \times \bm \Omega))},  \label{S1} \\
(\sqrt{\omega} \cdot \dot{{\bm x}})_i^\lambda&=&( 1{- \frac{1}{2}  (\bm \Omega \times \bm x)^2}){\bm \nu}_{0i}^\lambda + \bm e_i \times \bm b_i^\lambda \nonumber \\
&&
+   \hat{\bm p}  \cdot \bm b_i^\lambda(q_i \bm B+ 2p \bm \Omega) ( 1{- \frac{1}{2}  (\bm \Omega \times \bm x)^2)
	+ \hat{\bm p} \cdot \bm b_i^\lambda  [ (\bm x \times \bm \Omega) \times \bm e_i ] }
,\label{smlx} \label{S2}\\
(\sqrt{\omega} \cdot \dot{{\bm p}})_i^\lambda &=& \bm e_i+ {\bm \nu}_{0i}^\lambda \times (q_i\bm{B} + 2{\cal E}_i^\lambda \bm{\Omega})( 1{- \frac{1}{2}  (\bm \Omega \times \bm x)^2}) \nonumber \\
&&+  \bm b_i^\lambda (\bm{e}_i \cdot (q \bm{B} + 2p \bm{\Omega})){- [ (\bm x \times \bm \Omega) \times \bm e_i ] \times {\bm \nu}_{0i}^\lambda }
.  \label{smlp} \label{S3} 
\end{eqnarray}
The fictitious centrifugal force ${\cal E}_i^\lambda  \bm \Omega \times (\bm \Omega \times \bm x)$ and the Lorentz force due to the electric field $\bm E^\prime$   are unified as
\begin{equation}
\label{e}
\bm e_i =  q_i\bm E + (\bm\Omega\times \bm x)\times (q_i\bm B +{\cal E}_i^\lambda\bm\Omega).
\end{equation}
We define the chiral particle (antiparticle) number and current densities as
\begin{eqnarray}
n_i^\lambda  & = &  \int \frac{d^3p}{(2\pi\hbar)^3} (\sqrt{\omega})_i^\lambda  f_\lambda^i ,\label{nil} \\
\bm j_i^\lambda & = & \int \frac{d^3p}{(2\pi\hbar)^3} (\sqrt{\omega} \cdot \dot{\bm x})_i^\lambda f_\lambda^i  + \bm j_{\ssM i }^{\lambda} , \label{jil}
\end{eqnarray}
where the divergenceless  magnetization current \cite{syD,cssyy,cipy} is  
\begin{equation}
\label{curl}
 \bm j_{\ssM i}^{\lambda}=\bm \nabla \times  \int \frac{d^3p}{(2\pi\hbar)^3}    {\cal E}_i^\lambda  \bm b_i^\lambda f_\lambda^i .
\end{equation}
The magnetization current (\ref{curl}), which has also been derived  from the quantum field theory \cite{hpy},  is important for systems which are not in equilibrium \cite{ksy}. 

Distribution functions are defined to satisfy the Boltzmann transport equation, 
\begin{equation}
\frac {df_\lambda^i}{dt}=\frac{\partial f_\lambda^i}{\partial t}+\frac{\partial f_\lambda^i}{\partial \bm{x}}\cdot \dot{\bm{x}}+\frac{\partial f_\lambda^i}{\partial \bm{p}}\cdot \dot{\bm{p}}= (I_{\mathrm{coll}})^i_\lambda ,
\label{collboltzmann}
\end{equation}
where the collisions should conserve the number of particles:
\begin{equation}
\label{col-con}
\int \frac{d^3p}{(2\pi\hbar)^3} (\sqrt{\omega})_i^\lambda (I_{\mathrm{coll}})^i_\lambda=0.
\end{equation}
By making use of (\ref{col-con}) one can show that the 4-divergence of the  4-current $(n_i^\lambda,\bm j_i^\lambda )$ yields 
the continuity equation with source:
\begin{equation} 
\label{CE}
\frac{\partial n_i^\lambda}{\partial t} + \bm {\nabla} \cdot \bm j_i^\lambda = \int \frac{d^3p}{(2\pi\hbar)^3}  \left[ (\frac{1}{2}d\omega_i^\lambda \wedge \omega_i^\lambda \wedge \omega_i^\lambda)_{\ssV\ssM} f_i^\lambda  \right] .
\end{equation}
By an explicit calculation one finds 
\begin{eqnarray}
(\frac{1}{2}d\omega_i^\lambda \wedge \omega_i^\lambda \wedge \omega_i^\lambda)_{\ssV\ssM} &=& 2\pi \hbar \lambda q^2\delta(\bm p)\bm{{ E}}\cdot \bm{B} ,\label{vma}
\end{eqnarray}
where the subscript $VM$ denotes that the canonical volume form $dV\wedge dt$ is factored out and the  Maxwell equations, (\ref{Max}), have been employed. Therefore the continuity equation (\ref{CE})  is written as
\begin{equation} 
\label{ceqD0}
\frac{\partial n_i^\lambda}{\partial t} + \bm {\nabla} \cdot \bm j_i^\lambda = \frac{\lambda q^2}{(2\pi\hbar)^2}  \bm{{ E}}\cdot \bm{B} \ f_i^\lambda|_{\bm p =0}   .
\end{equation}
Note that we keep terms up to first order in $\hbar,$ so that only the $\hbar$ independent part of $ f_i^\lambda$ contributes  in the right-hand side of (\ref{ceqD0}).

We  ignore the side-jump effects \cite{cssyy, dehz} and consider  a Lorentz scalar $ f_i^\lambda.$ Thus, 
the 4-current density $(n_i^\lambda , \bm j_i^\lambda),$ 	does not transform as a Lorentz vector. Although the magnetization term is needed to define a covariant current, it is not sufficient in the presence of  collisions as discussed  in \cite{css} within a semiclassical approach and  in \cite{hpy} by means of quantum field theory.

We will consider the collisions  within the relaxation time approach. The relaxation time, $\tau,$ can be defined by specifying the scattering process of chiral particles and in general may depend on the velocity. However,  an acceptable approximation is to 
 let  the relaxation time be constant as we do here.
 The consistency condition (\ref{col-con})
can be guaranteed to be satisfied by choosing the collision term as \cite{mtr}
\begin{equation}
\label{bgk}
I_{\mathrm{coll}}= -\frac{1}{\tau} (f-\frac{n }{n_{eq}}f_{eq}),
\end{equation}
where the density  $n_{eq}$ is given by the equilibrium distribution function $f_{eq} .$ 
However,  instead of (\ref{bgk}), we  adopt the definition
\begin{equation}
\label{rtc}
(I_{\mathrm{coll}})^i_\lambda =-\frac{1}{\tau} (f^i_\ssl-f^{i}_{eq \ssl}).
\end{equation}
Thus  the condition (\ref{col-con})  leads to 
\begin{equation}
\label{con}
\int \frac{d^3p}{(2\pi\hbar)^3}  
 (\sqrt{\omega})_i^\lambda (f^i_\ssl-f^{i}_{eq \ssl)}) =0.
\end{equation}
Actually, the time  evolution equations of  inhomogeneous  chemical potentials  consistent with the continuity equation (\ref{ceqD0}),  are provided by the  condition (\ref{con}) as it will be shown explicitly.
Although, due to this condition
 only $f^{i}_{eq \ssl}$ contributes to the number density,
 to calculate the current densities and the evolution equations of chemical potentials one should be equipped with  the distribution functions obeying the Boltzmann equation (\ref{collboltzmann}).

\section{Solution of the Boltzmann transport equation}
\label{SSBE}

At first sight it might appear that the equilibrium distribution function should be  given by the  Lorentz scalar Fermi-Dirac distribution 
\begin{equation}
\label{eqf}
f^{i}_{eq \ssl} =\frac{1}{e^{[ p_\mu v^\mu-\mathrm{sign}  (q_i)\mu_\ssl ]/T} +1}  ,
\end{equation} 
where the  momentum and velocity 4-vectors  are  $p_\mu =({\cal{E}}_i^\lambda , \bm p) $ and $v_\mu=(1,\bm \Omega \times \bm x).$ However,   both the angular velocity $\bm \Omega$ and the linear velocity due to rotation 
$\bm v =\bm \Omega \times \bm x,$ have already been incorporated into the semiclassical approach as it is demonstrated  in Appendix A. There, we further showed that  in statistical equilibrium the calculations of number and current densities yield the same answer either if one  keeps the fully fledged dispersion relation (\ref{disp}) with (\ref{eqf}) or  approximates   them by ignoring their $\hbar$ dependent terms.
Therefore,   we drop the $\bm p \cdot (\Omega \times \bm x)$ term in (\ref{eqf}) and set $p_0=p:$
\begin{equation}
f^{i}_{eq \ssl} =f^{0i}_\ssl =\frac{1}{e^{[p-\mathrm{sign}  (q_i)\mu_\ssl ]/T} +1} . \label{feq0}
\end{equation}
We deal with ${\cal{E}}_i^\lambda =p,$ so that  the canonical velocity is 
${\bm \nu}_{0i}^\lambda=\hat{\bm p}.$ 
 Thus the kinetic equation which should be solved is 
\begin{eqnarray}
(1+ \bm{b}_i^\lambda \cdot \bm{A}_i { -\hat{\bm p} \cdot (\bm x \times \bm \Omega) -   \hat{\bm p}  \cdot \bm b_i^\lambda(q_i \bm B \cdot (\bm x \times \bm \Omega))}) \frac{\partial f^i_\lambda}{\partial t} \nonumber\\ +
\{ \bm{e}_i+\hat{\bm p} \times \bm{A}_i ( 1{- \frac{1}{2}  (\bm \Omega \times \bm x)^2}) + \bm{b}_i^\lambda (\bm{e}_i \cdot \bm{A}_i) {- [ (\bm x \times \bm \Omega) \times \bm e_i ] \times \hat{\bm p} }  \} \cdot \frac{\partial f^i_\lambda}{\partial \bm p}  \nonumber\\ +
\{ \hat{\bm p} ( 1{- \frac{1}{2}  (\bm \Omega \times \bm x)^2}) + \bm{e}_i \times \bm{b}_i^\lambda + \bm{A}_i (\bm{b}_i^\lambda \cdot \hat{\bm p})( 1{- \frac{1}{2}  (\bm \Omega \times \bm x)^2) }  \nonumber\\
+{( \hat{\bm p} \cdot \bm b_i^\lambda ) [ (\bm x \times \bm \Omega) \times \bm e_i ] } \} \cdot \frac{\partial f^i_\lambda}{\partial  \bm x}  = \nonumber\\  
-\frac{1}{\tau} (1+ \bm{b}_i^\lambda \cdot \bm{A}_i { - \hat{\bm p} \cdot (\bm x \times \bm \Omega) -  \hat{\bm p}  \cdot \bm b_i^\lambda(q_i \bm B \cdot (\bm x \times \bm \Omega))})  (f^i_\lambda-f^{0i}_\lambda),
\label{cbe}
\end{eqnarray} 
where $\bm A_i\equiv q_i  \bm{B}+2 p \bm{\Omega}.$ 
The derivatives of equilibrium distribution function  will show up  extensively in the calculations, so that we introduce the short-handed notation  $D^i_\ssl \equiv - T\partial f^{0i}_\lambda /\partial p =\mathrm{sign}  (q_i) T\partial f^{0i}_\ssl /\partial \mu_\ssl .$

To simplify  the notation let us only examine  the right-handed particles by suppressing the indices. Its generalization to antiparticles and  the left-handed  particles is  straightforward. To solve the kinetic equation  (\ref{cbe}) there are some  approximations  in  order. 
Inspired by the solution of the Boltzmann equation for  nonrelativistic electrons in the presence of electromagnetic fields \cite{anselm},  
we start by calculating the distribution function linear in   $\bm e $ and the derivatives of the chemical potential.  The former is equivalent to consider  first order terms in $\bm E$ and  $(\bm x \times \bm \Omega) .$ Note that  derivatives of $\bm{e}$ are regarded  to be second order for being able to solve the kinetic equation (\ref{cbe}).  Although this is the standard procedure \cite{anselm},  it may   be misleading  in some cases even for the  electric field $\bm E$ due to the Maxwell equations (\ref{Max}).  We will see that to overcome some inconsistencies caused by this approximation, in some cases we need to  choose $\bm E$  and $\bm \Omega$ to be mutually perpendicular.
In the first order we set 
$ f = f^{0} +f^{1} $ and solve  (\ref{cbe}) for $f^1.$ If we retain only the terms depending on $f^0$ on the left-hand side of
(\ref{cbe})  we cannot take into account the $\bm A$ field, so that we should also maintain the
$\partial f^1/\partial \bm p$ dependent terms. By examining the Boltzmann equation one can observe that $f^1$  can be taken in the form
$$
f^1= -\frac{\partial f^0}{\partial p} ( \bm{\chi} \cdot \bm{p})-\tau \frac{\partial f^0}{\partial \mu}
\frac{\partial \mu }{\partial t}.
$$
Now our task is to find the  functional $\bm \chi.$  By ignoring the terms higher than the first order in $\bm e$ and  $\partial \mu /\partial \bm x ,$
we acquire
\begin{equation}
\label{bolf1}
-\frac{\partial f^0}{\partial p}\bm{\hat p} \cdot [\bm e_\mu +(\bm e_\mu \cdot \bm A) \bm b-\bm A \times \bm \chi]] = \frac{1}{\tau} (1+ \bm{b} \cdot \bm{A})\frac{\partial f^0}{\partial p} \bm p \cdot\bm \chi .
\end{equation}
We defined  $\bm e_\mu$   by substituting   $\bm E$  with 
\begin{equation}
\bm E_\mu=  \bm E  - \frac{1}{q}\frac{\partial \mu}{\partial \bm x} ,
\label{Emu}
\end{equation}
in (\ref{e}).
By expanding $\bm\chi$ up to  first order in the Planck constant as
$
\bm\chi= \bm \chi^0 + \bm \chi^1,
$
where $\bm \chi^0$ depends only on the magnitude of momentum, $p,$ 
(\ref{bolf1}) yields the  following coupled equations 
\begin{eqnarray}
\bm{e_\mu} - \bm{A} \times \bm \chi^0 & = & \frac{p}{\tau} \bm \chi^0 \label{chizero} , \\
\bm{\hat p} \cdot \left[\bm b (\bm{e_\mu} \cdot \bm{A}) - \bm{A} \times \bm \chi^1\right]& = & \frac{\bm p}{\tau}  \cdot \left[ \hbar \bm \chi^1 + (\bm{b} \cdot \bm{A}) \bm \chi^0\right]\label{chione}.
\end{eqnarray}
Equation (\ref{chizero}) can be solved as
\begin{equation}
\bm \chi^0= \frac{g \bm{e_\mu}  - g^2 \bm{A} \times \bm e_\mu + g^3 \bm{A} (\bm{e_\mu} \cdot \bm{A}) }{1+g^2 A^2},
\label{chi0}
\end{equation}
where $g\equiv \tau \slash p.$ 
By observing  that the Berry curvature dependence of  $\bm \chi^1 $ should be in the form $\bm \chi^1 = \bm \chi^1(\bm b, \bm b \cdot \bm A),$
one can show that 
\begin{equation}
\bm{\chi }^1= g (\bm{e_\mu} \cdot \bm{A})\bm b -(\bm{b} \cdot \bm{A}) \frac{ \bm \chi^0-g \bm{A} \times \bm \chi^0+g^2 \bm A (\bm A \cdot \bm \chi^0) }{1+g^2 A^2}
\end{equation}
solves  (\ref{chione}).
Hence,   at the first order in $\bm e_\mu$  the distribution function is established as 
\begin{eqnarray}
f^1=	\frac{\tau D_\ssl}{T}\Big[- \frac{\partial \mu}{\partial t}+  \hat{\bm{p}}\cdot  \left( 
\frac{ \bm{e_\mu} - g \bm{A} \times \bm e_\mu + g^2 \bm{A} (\bm{e_\mu} \cdot \bm{A})(1-{\bm b \cdot \bm A})}{1+g^2 A^2}+ {\bm b (\bm{e_\mu} \cdot \bm{A})} \right) \nonumber \\
-	\hat{\bm{p}}\cdot\frac{ (1-g^2A^2)\bm{e_\mu}  -2 g \bm{A} \times \bm e_\mu +2 g^2 \bm{A} (\bm{e_\mu} \cdot \bm{A}) }{(1+g^2 A^2)^2} (\bm{b} \cdot \bm{A}) \Big] .\label{f1o}
\end{eqnarray}
Computation of the condition (\ref{con}) for $f^1$ reveals that
$\partial \mu /\partial t$ is at the order of $\hbar.$ 

It is  awkward to calculate the distribution function beyond the first order. In spite of this, we would like to obtain the second order solution  $f^{2}$ satisfying
\begin{eqnarray}
\frac{ D}{T} (1+ \bm{b} \cdot \bm{A}{ - \hat{\bm p} \cdot (\bm x \times \bm \Omega) - (  \hat{\bm p}  \cdot \bm b)(q \bm B \cdot (\bm x \times \bm \Omega))})\frac{\partial}{\partial t} \left(-\tau\frac{\partial \mu }{\partial t} +  \bm \chi \cdot \bm p\right) && \nonumber \\
  + \frac{ D}{T^2} (1-2 f^0) \left(-\tau\frac{\partial \mu }{\partial t} +  \bm \chi \cdot \bm p\right) [ \hat{\bm p} \cdot \bm e_\mu + ( \hat{\bm p} \cdot \bm b) \ (\bm A \cdot \bm e) ] &&\nonumber\\
+ \frac{ D}{T}  \frac{\partial \left(  \bm \chi \cdot \bm p\right)}{\partial \bm p}  \cdot [\bm e + \bm b (\bm A \cdot \bm e) ]
+ \frac{ D}{T} (\bm e \times \bm b) \cdot  \frac{\partial \mu}{\partial \bm x} - \frac{ D}{T}  \hat{\bm p} \cdot  \frac{\partial}{\partial \bm x}\left(-\tau\frac{\partial \mu }{\partial t} +  \bm \chi \cdot \bm p\right)  =&& \nonumber\\
{ (f^1/ \tau) \left[ \hat{\bm p} \cdot (\bm x \times \bm \Omega) +  \hat{\bm p}  \cdot \bm b (q\bm B \cdot (\bm x \times \bm \Omega))\right] } -\frac{(1+ \bm{b} \cdot \bm{A})}{\tau}\ f^{2} , && \nonumber
\end{eqnarray} 
which is required by (\ref{cbe}).
It is solved by
\begin{eqnarray}
f^{2}
&=& -\frac{\tau D}{T}  \frac{\partial}{\partial t} \left(-\tau\frac{\partial \mu }{\partial t} +  \bm \chi \cdot \bm p\right)\nonumber\\ 
	&&+ (1- \bm{b} \cdot \bm{A}) \left(  - \frac{\tau D}{T^2} \left( (1-2 f^0) \ (\bm \chi^0 \cdot \bm p) ( \hat{\bm p} \cdot \bm e_\mu +  T \  \frac{\partial (\bm \chi^0  \cdot \bm p)}{\partial \bm p} \cdot \bm e + T \  \hat{\bm p} \cdot  \frac{\partial (\bm \chi^0 \cdot \bm p)}{\partial \bm x} \right) \right)\nonumber\\
&& -\frac{\tau D}{T^2} (1-2 f^0) \left[ \left(-\tau\frac{\partial \mu }{\partial t} +  \bm \chi^1 \cdot \bm p\right) ( \hat{\bm p} \cdot \bm e_\mu + (\hat{\bm p} \cdot \bm b) \ (\bm A \cdot \bm e) (\bm \chi^0 \cdot \bm p)   \right]\nonumber\\
&&- \frac{\tau D}{T}  \left( \frac{\partial (\bm \chi^1 \cdot \bm p)}{\partial \bm p} \cdot \bm e + \frac{\partial (\bm \chi^0  \cdot \bm p)}{\partial \bm p} \cdot \bm b (\bm A \cdot \bm e)  \right)  \nonumber\\
&&- \frac{\tau D}{T} (\bm e \times \bm b) \cdot  \frac{\partial \mu}{\partial \bm x} - \frac{\tau D_\ssl}{T}   \hat{\bm p} \cdot  \frac{\partial}{\partial \bm x} \left(-\tau\frac{\partial \mu }{\partial t} +  \bm \chi^1 \cdot \bm p\right) \nonumber \\
&& {- f^1 \left[ \hat{\bm p} \cdot (\bm x \times \bm \Omega) +  \hat{\bm p}  \cdot \bm b (q\bm B \cdot (\bm x \times \bm \Omega))\right] } . \label{f2o}
\end{eqnarray}
Even in this closed form it is composed of several terms. Thus, we will 
 take into account some specific cases. We would like to recall that we dealt with the right-handed fermions which can easily be generalized to the left-handed  fermions and antiparticles. 

\section{Particle number and current densities}
\label{pnc}

The nonequilibrium distribution functions are subject to the condition (\ref{con}), hence the number density involves only the equilibrium distribution function, $f_0.$ By inserting the phase space measure  (\ref{S1}) into the definition (\ref{nil}) and discarding the terms whose integrals  clearly  vanish,  the chiral number density  turns out to be
\begin{equation}
n_i^\lambda = \int \frac{d^3p}{(2\pi\hbar)^3} [1-   \hat{\bm p}  \cdot \bm b_i^\lambda(q_i \bm B \cdot (\bm x \times \bm \Omega))]f^{0i}_\lambda . \label{nLL}
\end{equation}
On the other hand by employing the velocity  (\ref{S2}) in  (\ref{jil}),  the chiral current density which is given with the full distribution function becomes
\begin{eqnarray}
	\bm j_i^\lambda  &=&   \int \frac{d^3p}{(2\pi\hbar)^3} \{ \hat{\bm p} ( 1{- \frac{1}{2}  (\bm \Omega \times \bm x)^2}) + \bm{e}_i \times \bm{b}_i^\lambda  \nonumber\\
	&+&\bm{A}_i (\bm{b}_i^\lambda \cdot \hat{\bm p})( 1{- \frac{1}{2}  (\bm \Omega \times \bm x)^2) } +{( \hat{\bm p} \cdot \bm b_i^\lambda ) [ (\bm x \times \bm \Omega) \times \bm e_i ] } \} f_\lambda^i . \label{jLL}
\end{eqnarray}
By means of the Fermi-Dirac integrals provided in Appendix A, one can readily calculate  the chiral number  density, (\ref{nLL}), as
\begin{equation}
n_\ssl = \frac{\mu_\ssl }{6 \pi^2\hbar^3} (\mu_\ssl ^2 + \pi^2 T^2)-  \frac{q\mu_\ssl}{4 \pi^2\hbar^2}  \bm B \cdot (\bm x \times \bm \Omega). \label{nll}
\end{equation}
In the calculation of nonequilibrium distribution function we treat the linear velocity $\bm v = \bm \Omega \times \bm x$  as perturbation, moreover, the last term is at the order of $\hbar$ with respect to the first term. Hence, we approximate the  chiral number density as
$$
n_\ssl = \frac{\mu_\ssl }{6 \pi^2\hbar^3} (\mu_\ssl ^2 + \pi^2 T^2).
$$
We sum the  particle and antiparticle contributions to get the number and current densities of chirality $\lambda:$ 
\begin{equation}
	n_\lambda =\sum \limits_i \mathrm{sign}(q_i)n_i^\lambda,  \ \ \ \ 
	j_\lambda=\sum \limits_i \mathrm{sign}(q_i)\bm j_i^\lambda . \nonumber
\end{equation}
The total and axial  number densities are defined as 
\begin{equation}
	n = \sum_\ssl n_\ssl  , \ \ \ \  n_5= \sum_\ssl  \lambda \ n_\ssl . \nonumber
\end{equation}
The vector and axial current densities are  defined similarly:
\begin{equation}
\bm j = \sum_\ssl \bm j_\ssl , \ \ \ \ \bm j_5= \sum_\ssl  \lambda \ \bm j_\ssl .
\label{eac}
\end{equation}
Note that  the electric charge and current densities are given by $qn$ and $q\bm j.$ 

In terms of  the total chemical potential $\mu =\frac{1}{2}(\mu_\ssR +\mu_\ssL)$ and the chiral chemical potential $\mu_5 =\frac{1}{2}(\mu_\ssR -\mu_\ssL)$ we find
\begin{eqnarray}
n &=& \frac{\mu}{3 \pi^2\hbar^3} (\mu^2 + 3 {\mu_5}^2 + \pi^2 T^2), \label{nvv}\\
n_5 &=& \frac{\mu_5}{3 \pi^2\hbar^3} (3 \mu^2 + {\mu_5}^2  + \pi^2 T^2 ). \label{n55}
\end{eqnarray}
Let us introduce the charge and ``axial charge'' susceptibilities:
\begin{equation}
\label{susp}
\chi\equiv \frac{\partial n}{\partial \mu}  = \frac{ \mu^2 +  {\mu_5}^2}{\pi^2\hbar^3} + \frac{T^2}{3 \hbar^3};\ \ \ \chi_5 \equiv \frac{\partial n}{\partial \mu_5} =\frac{2 \mu \mu_5}{ \pi^2 \hbar^3} .
\end{equation}
Observe that they can also be written as $\chi=\partial n_5\slash \partial \mu_5,\ \chi_5= \partial n_5/\partial \mu.$

The calculation of currents by making  use of the fully fledged distribution functions  on general grounds is formidable
even if we consider only contributions arising from $f_1.$ Hence we deal with some specific cases. The calculations of  currents should  be supplemented by the evolution equations of chemical potentials  in  time, delivered by the consistency condition (\ref{con}).

\subsection{ Current densities for $ T=0$}
\label{st0}
Although we  mainly deal with hot plasma, to get an idea about the structure of currents  we first would like to  present the $T=0$ case.  At zero temperature antiparticles do not contribute to the number and current densities. The  current density arising from the Fermi-Dirac distribution function, (\ref{feq0}), without any approximation is
\begin{eqnarray}
\bm j_\ssl(f^0)_{T=0}  &=& \frac{\lambda  \mu_\ssl}{4\pi^2\hbar^2}(q\bm B +\mu_\ssl \bm \Omega ) 
\nonumber\\
&& +\frac{\lambda  \mu_\ssl}{4\pi^2\hbar^2}
\left[q (\bm x \times \bm \Omega)\times \bm E -  q (\bm \Omega \times \bm x) (\bm \Omega \times \bm x)\cdot \bm B- \frac{\mu_\ssl}{2}\bm \Omega (\bm \Omega \times \bm x)^2
\right]. \label{ft0}
\end{eqnarray}
However  $f^1_\ssl $ is obtained 
by keeping terms  up to linear order in $\bm E,\ (\bm\Omega\times \bm x) $ and derivatives. Hence it gives rise to the current density
\begin{eqnarray}
\bm j_\ssl(f^1)_{T=0}  &=& \frac{ \tau \mu_\ssl}{6\pi^2\hbar^3} \Big[ \frac{3\lambda \hbar  }{2\mu_\ssl} \frac{\partial \mu_\ssl }{\partial t}(q\bm B +\mu_\ssl\bm \Omega ) 
\label{ft1}\\
&&+\frac{\mu_\ssl^2  \tilde{\bm   e}_{\mu_\ssl} - \tau \mu_\ssl (q\bm B +2 \mu_\ssl \bm \Omega ) \times  \tilde{\bm   e}_{\mu_\ssl} + \tau^2(q\bm B +2\mu_\ssl\bm \Omega ) [ \tilde{\bm   e}_{\mu_\ssl}  \cdot (q\bm B+2\mu_\ssl\bm \Omega )]}{\mu_\ssl+ \tau^2 (q\bm B +2\mu_\ssl\bm \Omega )^2} \Big],
\nonumber
\end{eqnarray}
where we introduced the short-handed notation 
$$
\tilde{\bm   e}_{\mu_\ssl} =  q\bm E-\bm \nabla \mu_\ssl + (\bm\Omega\times \bm x)\times (q\bm B +\mu_\ssl\bm\Omega).
$$
The divergence of  $\bm j_\ssl(f^1)$ yields  the derivatives of  $ \tilde{\bm   e}_{\mu_\ssl},$ which are considered as second order as well as 
second order terms in the derivatives of $\mu_\ssl,$  so that we get \mbox{$\bm \nabla \cdot \bm j_\ssl(f^1)_{T=0} \approx 0.$ }
On the other hand for $f^1_\ssl $ the consistency condition (\ref{con}) leads to
\begin{equation}
\label{ct1}
\mu_\ssl^2 \frac{\partial \mu_\ssl }{\partial t} -\frac{\lambda \hbar}{2} ( q\bm  E -\bm \nabla \mu_\ssl )\cdot (q\bm B  +2 \mu_\ssl \bm \Omega ) -\frac{\lambda \hbar q\mu_\ssl}{2}(\bm \Omega \times \bm x)\cdot (\bm B \times \bm \Omega)=0.
\end{equation}
If we consider the current density to be the sum of (\ref{ft0}) and (\ref{ft1}), the continuity equation 
$$
\frac{\partial n_\lambda}{\partial t}+\bm \nabla \cdot \bm j_\lambda = \frac{\hbar  \lambda q^2}{4\pi^2 \hbar^3} \bm E \cdot \bm B 
$$
is satisfied as far as  the time evolution equation (\ref{ct1}) is fulfilled. However, we should approximate also the current due to $f^0_\ssl$ by ignoring the terms in second line of (\ref{ft0}). But now to satisfy the continuity equation  we should deal with the fields satisfying $\bm E \cdot \bm \Omega =0, $ and $\bm \Omega \times \bm B =0.$ This inconvenience disappears when we include the second order terms.  

Plugging (\ref{ft0}) and (\ref{ft1})  into  (\ref{eac}) will lead to the vector 
and  axial current densities 
at zero temperature.

\section{Current densities for  $\bm B=0$}
\label{b0}
The main objective of this work is to reveal rotation dependent nonlinear phenomena in chiral plasma. In accord with this scope it is appropriate to 
consider the current densities  generated by the rotation of coordinate frame and  electric field, in the absence of magnetic field. Thus, throughout this section we set $\bm B=0$ in the expressions which have been obtained in the preceding sections. 

The current density which results from the Fermi-Dirac distribution, (\ref{feq0}),  can readily be obtained by ignoring the centrifugal terms, 
as
\begin{eqnarray}
\bm j_\ssl (f^0)  &=&\frac{\lambda (3 {\mu_\ssl}^2 + \pi^2 T^2)}{12 \pi^2\hbar^2 }  \bm \Omega + \frac{\lambda q \mu_\ssl}{4 \pi^2\hbar^2 }   (\bm x \times \bm{\Omega})  \times \bm E . \label{jlf0}
\end{eqnarray}
For vanishing magnetic field the denominators of (\ref{f1o}) and (\ref{f2o}) are independent of the energy, so that we can perform the phase space integrals by making use of the Fermi-Dirac integrals given  in Appendix A. Actually the chiral current density arising from   $f^1_\ssl$ is calculated as
\begin{eqnarray}
\bm j_\ssl (f^1)  =&& -\frac{\lambda\tau \mu_\ssl }{2 \pi^2 \hbar^2}\frac{\partial \mu_\ssl}{\partial t} \bm \Omega \nonumber \\
&&+  \frac{\tau q\left(3 {\mu_\ssl}^2 + \pi^2 T^2\right)}{18 \pi^2\hbar^3(1+4 \tau^2 \Omega^2) }\left[\bm E_{\mu_\ssl}  -2 \tau\bm{\Omega} \times  \bm E_{\mu_\ssl} +4 \tau^2 \bm{\Omega} (\bm E_{\mu_\ssl}\cdot \bm{\Omega})\right].  \label{jlf1}
\end{eqnarray}
At this order the consistency condition (\ref{con})    yields 
\begin{equation}
\label{coc1}
 - \frac{(\pi^2 T^2+ 3 \mu_\lambda^2)}{6 \pi^2\hbar^3} \frac{\partial \mu_\lambda }{\partial t} + \frac{\lambda  \tau \mu_\lambda }{2 \pi^2\hbar^2}  \left(q\bm E \cdot \bm \Omega  -\bm \nabla  \mu_\ssl  \cdot \bm \Omega \right)	=0. 
\end{equation}
Hence,  $\partial \mu_\ssl/\partial t$  is at  the order of $\hbar$ and  linear  in $\bm  E$  and $\bm \nabla \mu_\ssl .$ 

To fully attain  the second order current densities  we need to take the trouble of calculating a large number of terms.  The  ones originating from the centrifugal force related terms of $f^2$ are particularly lengthy, so that we ignore them here. Nevertheless for completeness they are furnished in  Appendix B.  After some calculations the chiral currents which are second order in $\bm  E,$  $ (\bm \Omega \times \bm x )$ and the derivatives are obtained as
\begin{eqnarray}
	\bm j_\ssl ^{(2)}  &=	&
	-q \tau^2 \frac{(3 {\mu_\ssl}^2 + \pi^2 T^2)}{18 \pi^2\hbar^3 (1+4 \tau^2 \Omega^2) }\left[ \frac{\partial \bm E}{\partial t}  -2 \tau \bm{\Omega} \times   \frac{\partial \bm E}{\partial t} +4 \tau^2 \bm{\Omega} (  \frac{\partial \bm E}{\partial t}\cdot \bm{\Omega})
	\right]	\nonumber\\
	&&+  \frac{ \lambda \tau^2}{\pi^2 \hbar^2 (1+4\tau^2 \Omega^2)^2}\bm \Omega 
	\bigg\{ -\frac{(1+16 \tau^2 \Omega^2) }{15}\bm E^2 
	+\frac{(1-12 \tau^2 \Omega^2)}{30} \bm \nabla\mu_\ssl \cdot \bm E \nonumber \\ 
	&&
	+\frac{(1-20 \tau^2 \Omega^2) }{30}(\bm \nabla \mu_\ssl)^2  + \frac{4\tau^2}{15} \left[(8+23 \tau^2 \Omega^2) (\bm E\cdot \bm \Omega)^2 
 +(8+14 \tau^2 \Omega^2) (\bm \Omega \cdot \bm \nabla \mu_\ssl)^2 \right] \nonumber\\
 &&-  (16+37 \tau^2 \Omega^2)  (\bm E \cdot \bm \Omega) \bm \Omega \cdot \bm \nabla \mu_\ssl- \frac{(7-32 \tau^2 \Omega^2)}{15}\tau \bm E \cdot (\bm \Omega \times \bm \nabla \mu_\ssl)
   \bigg\}  \nonumber\\
 &&- \frac{4 \tau^2 \lambda}{15 \pi^2 \hbar^2} \bm E (\bm E_{\mu_\ssl}\cdot \bm \Omega) + \frac{ \tau^2 \lambda}{\pi^2 \hbar^2(1+4 \tau^2 \Omega^2)} \bm E_{\mu_\ssl} \bigg\{  \frac{\bm E_{\mu_\ssl}\cdot \bm \Omega}{30}(7+8 \tau^2 \Omega^2+16 \tau^4 \Omega^4)\nonumber\\
 	&&-\frac{\bm E \cdot \bm \Omega}{3}(2\tau^2 \Omega^2)\bigg\}+\frac{ \tau^3 \lambda }{\pi^2 \hbar^2(1+4 \tau^2 \Omega^2)^2}(\bm \Omega \times \bm E_{\mu_\ssl}) \bigg\{ \frac{3+4 \tau^2 \Omega^2}{5} \bm E_{\mu_\ssl}\cdot \bm \Omega\nonumber\\
 	&&- \frac{6+8 \tau^2 \Omega^2}{15} \bm E \cdot \bm \Omega\bigg\}+\frac{ \lambda \tau^3\Omega^2 }{3 \pi^2 \hbar^2(1+4 \tau^2 \Omega^2)}\bm E  \times \bm \nabla  \mu_\ssl \nonumber\\
 	 &&+ \frac{\lambda \tau \mu_\ssl}{30 \pi^2 \hbar^2(1+4\tau^2 \Omega^2)^2}\bm \Omega \bigg\{ \bm E_{\mu_\ssl} \cdot (\bm x \times \bm \Omega)(5+4\tau^2 \Omega^2)+4\tau \Omega^2 (\bm x \cdot \bm E_{\mu_\ssl})
 	 \nonumber\\
 	 &&-4\tau^2(4+8 \tau^2 \Omega^2)(\bm \Omega \cdot \bm x)(\bm \Omega \cdot \bm E_{\mu_\ssl}) \bigg\}\nonumber\\
 	  &&-\frac{\tau \lambda \mu_\ssl}{15 \pi^2 \hbar^2} (\bm x \times \bm \Omega) (\bm E_{\mu_\ssl}\cdot \bm \Omega)+ \frac{\lambda q^2 \tau^3}{3 \pi^2 \hbar^2(1+4 \tau^2 \Omega^2)}(\bm E \times \bm \Omega)(\bm \Omega \cdot \bm  E_{\mu_\ssl})\nonumber\\
 	  &&- \frac{\lambda q^2 \tau^2 }{6 \pi^2 \hbar^2 (1+4 \tau^2 \Omega^2)} \bm E \times (\bm \Omega \times \bm  E_{\mu_\ssl} )\nonumber\\   
 	  &&-  \frac{\tau^2 \lambda}{10 \pi^2 \hbar^2} \mu_\ssl \bm \Omega (\bm \nabla \cdot \bm K)+\frac{\tau^2 \lambda}{15 \pi^2 \hbar^2} \mu_\ssl \bm \Omega (\bm \nabla \cdot {\bm E_\mu}_\ssl)  + \frac{ \lambda \tau \mu_\ssl}{5\pi^2 \hbar^2}  \Psi_0[\bm E_{\mu_\ssl}] \nonumber\\ 
 	  &&+ \frac{ \lambda \tau}{12 \pi^2 \hbar^2 (1+4\tau^2 \Omega^2)^2} \big\{ \mu_\ssl  \bm \nabla \times \bm E_{\mu_\ssl} - \bm E_{\mu_\ssl} \times \bm \nabla \mu_\ssl -3 \tau \mu_\ssl  \bm \Omega (\bm \nabla \cdot  \bm E_{\mu_\ssl}) \nonumber\\
 	  &&+ 2 \tau \bm E_{\mu_\ssl} (\bm \nabla \mu_\ssl  \cdot \bm \Omega) + 4 \tau^2 (\bm \nabla \mu_\ssl  \times \bm \Omega) (\bm E_{\mu_\ssl} \cdot \bm \Omega)\big\} .
\label{jla2}
\end{eqnarray}
To simplify the presentation we introduced the functionals 
$$\bm K= \frac{  \tilde{\bm \chi}^0[\bm E_{\mu_\ssl}]-2 \tau \bm \Omega \times \bm  \tilde{\bm \chi}^0[\bm E_{\mu_\ssl}]+4 \tau^2 \bm \Omega (\bm \Omega \cdot \bm  \tilde{\bm \chi}^0[\bm E_{\mu_\ssl}]) }{1+4 \tau^2 \Omega^2},$$
$$ \Psi_0[\bm E_{\mu_\ssl}]= \frac{2 }{3} \bm{\Omega} (\bm \nabla \cdot  \tilde{\bm \chi}^0[\bm E_{\mu_\ssl}]) 
 -   \bm \nabla  (\bm \Omega \cdot  \tilde{ \bm \chi}^0[\bm E_{\mu_\ssl}]),  $$
in terms of  
\begin{equation}
 \tilde{\bm \chi}^0 [\bm E_{\mu_\ssl}]= \frac{\tau \bm E_{\mu_\ssl} - 2  \tau^2 \bm{\Omega} \times \bm E_{\mu_\ssl}  + 4 \tau^3 \bm{\Omega} (\bm E_{\mu_\ssl} \cdot \bm{\Omega}) }{1+4\tau^2 \Omega^2} .
 \label{cto}
\end{equation}
The last two lines in (\ref{jla2}) stem from the magnetization current (\ref{curl}). Here  we  do not explicitly  present  the consistency condition (\ref{con}) for  $f^2$  since it is too lengthy, though it would have guaranteed that
the continuity equation is satisfied:
$$
\frac{\partial n_\lambda}{\partial t}+\bm \nabla \cdot \bm j_\lambda = 0.
$$
Now, to establish the vector and axial  currents up to second order, it is sufficient to plug  (\ref{jlf1}) and (\ref{jla2}) in (\ref{eac}).
 
\subsection{The vector and axial  currents}
\label{vab01}

The current densities  up to  first order can be acquired  by employing  (\ref{jlf0}) and (\ref{jlf1}) in (\ref{eac}).  In fact
one can show that    the vector  current density is
\begin{eqnarray}
\bm j^{(1)}&= & \frac{\chi_5}{2} \bm \Omega   - \frac{\hbar \tau }{ 2}\frac{\partial \chi_5}{\partial t} \bm \Omega \nonumber\\
&& +\frac{\chi}{3} \frac{q \tau \bm E  -2 \tau^2 q\bm{\Omega} \times  \bm E +4 \tau^3 q\bm{\Omega} (   \bm E \cdot \bm{\Omega})}{1+4 \tau^2 \Omega^2} \nonumber\\ 
&&- \frac{\chi}{3}  \frac{\tau \bm \nabla \mu  -2 \tau^2 \bm{\Omega} \times  \bm \nabla \mu + 4 \tau^3 \bm{\Omega} (   \bm \nabla \mu \cdot \bm{\Omega})}{1+4 \tau^2 \Omega^2} \nonumber \\
&&- \frac{\chi_5}{3}  \frac{\tau \bm \nabla \mu_5  -2 \tau^2 \bm{\Omega} \times  \bm \nabla \mu_5 + 4 \tau^3 \bm{\Omega} (   \bm \nabla \mu_5 \cdot \bm{\Omega})}{1+4 \tau^2 \Omega^2} ,\label{jj11}
\end{eqnarray}
and the axial  current density  is
\begin{eqnarray}
\bm j_5^{(1)} &=& \frac{\chi}{2} \bm \Omega   - \frac{\hbar \tau }{ 2}\frac{\partial \chi}{\partial t} \bm \Omega  \nonumber\\
&&  +\frac{\chi_5}{3}  \frac{\tau \bm E  -2 \tau^2 \bm{\Omega} \times  \bm E +4 \tau^3 \bm{\Omega} (   \bm E\cdot \bm{\Omega})}{1+4 \tau^2 \Omega^2} \nonumber\\
&&- \frac{\chi_5}{3}  \frac{\tau \bm \nabla \mu  -2 \tau^2 \bm{\Omega} \times  \bm \nabla \mu + 4 \tau^3 \bm{\Omega} (   \bm \nabla \mu \cdot \bm{\Omega})}{1+4 \tau^2 \Omega^2} \nonumber \\
&&- \frac{\chi}{3}  \frac{\tau \bm \nabla \mu_5  -2 \tau^2 \bm{\Omega} \times  \bm \nabla \mu_5 + 4 \tau^3 \bm{\Omega} (  q \bm \nabla \mu_5 \cdot \bm{\Omega})}{1+4 \tau^2 \Omega^2}.\label{5511}
\end{eqnarray}
The charge and axial charge susceptibilities  $\chi$ and $\chi_5$ are given in (\ref{susp}).

Before proceed to present the higher order terms of  the current densities, we would like to elucidate  (\ref{jj11}) and  (\ref{5511}): The
first terms correspond to the chiral vortical and local polarization effects, respectively. The
second terms yield similar effects 
at the order of $\hbar,$  associated to the time derivatives of susceptibilities instead of the susceptibilities themselves.

To discuss the terms linear in the electric field $\bm E,$ first let it be parallel to 
$\bm \Omega.$ Now the second lines of (\ref{jj11}) and (\ref{5511}), respectively,  yield the Ohm's law,
$\bm j_{\mathrm{Ohm}} =(q\tau\chi/3)\bm E,$
and the electric separation effect, $ \bm j_{\mathrm{sep}}=(q\tau\chi_5/3)\bm E.$
Then let us focus on the case where $\bm E$ and $\bm \Omega$ are mutually perpendicular. For simplicity let us unify the vector and axial current densities linear in  $\bm E$ as follows
$$
\bm J\equiv \frac{q\alpha}{1+4 \tau^2 \Omega^2}  \left(  \tau \bm E - 2 \tau^2 \bm \Omega  \times \bm E \right),
$$ 
where $\alpha= \chi/3$ for the vector current density  and $\alpha=\chi_5/3$ for the  axial  current density.  We would like to show that it yields a Hall-like current. For this purpose 
let $\bm \Omega =\Omega \hat{z},$ and $\bm E =E_x\hat{x}+E_y\hat{y}.$ Thus $J_z$ vanishes and  the other components can be written as
\begin{eqnarray}
	J_x &=& a_1 E_x -a_2 E_y ,\label{Jx}\\
	J_y &=& a_2E_x +a_1 E_y , \label{Jy}
\end{eqnarray}
where  
$
	a_1=\frac{q \tau \alpha}{1+4 \tau^2 \Omega^2} , \ 
	a_2=\frac{2q \tau^2 \alpha \Omega}{1+4 \tau^2 \Omega^2}  . 
$
In the Hall effect there appears a current 
in the plane perpendicular to magnetic field which is also perpendicular to the applied electric field. Hence we further  set $J_y=0$ and express $E_x$ in terms of $E_y,$ which leads to
$$
J_x=- \sigma_{\ssH}^{(\Omega)}  E_y,
$$ 
with the conductivity
$$
\sigma^{(\Omega)}_\ssH =\left(\frac{a_1^2+a_2^2}{a_2}\right)=\frac{q\alpha }{2\Omega} . 
$$ 
It is  the Hall conductivity associated to the charge susceptibilities and to
 the effective magnetic field proportional to $2\Omega.$  It has been known that for the massive charge carriers there exist a rotational analogue of the Hall effect due to the Coriolis force \cite{ae}.
On the other hand we may express $E_y$ in terms of $E_x,$ which  yields
$$
J_x=q \tau \alpha  E_x.
$$ 
Hence the rotation of the coordinate frame does not alter  the Ohm's law and the charge separation effect.

By inspecting (\ref{jj11}) and  (\ref{5511}) one observes that the above discussions are valid also for
 the currents generated by the  gradients of chemical potentials by substituting $\bm E$ with $\bm\nabla \mu$ or $\bm\nabla \mu_5.$  

Although  we  postponed the discussion of the centrifugal force related terms until the Appendix B, the second order current densities, (\ref{jla2}), still possess too many terms. Thus  we present only few of them
explicitly. The remaining ones  can be read from (\ref{jla2}). The gradients of  chemical potentials behave like  the electric field. Thus, for simplicity we  deal with the current densities associated to $\bm E$  by setting  $\bm \nabla \mu_\ssl =0 .$ 
Let us first deal with the mutually perpendicular electric field and angular velocity: $\bm E\cdot \bm \Omega =0.$  The vector current density can be shown to become
\begin{eqnarray}
\bm j_\perp^{(2)}  &=& - \frac{\tau^2 q \chi }{3(1+4 \tau^2 \Omega^2)}\left(\frac{   \partial \bm E}{\partial t}  -2 \tau\bm{\Omega} \times   \frac{\partial \bm E}{\partial t} \right) - \frac{\mu_5}{ 2 \pi^2 \hbar^2 } (\bm x \times \bm{\Omega}  ) \times \bm E  
\nonumber\\
&&+ \frac{\tau^2}{5 \pi^2 \hbar^2}  \mu_5 \bm \Omega (\bm \nabla \cdot \bm E) \left( \frac{2}{3} - \frac{7}{30(1+4\tau^2 \Omega^2)}- \frac{1 -4 \tau^2 \Omega^2}{(1+4 \tau^2 \Omega^2)^2}  \right)
\nonumber\\
&&+ \frac{\tau \mu_5}{15 \pi^2 \hbar^2(1+4\tau^2 \Omega^2)^2}\bm \Omega \left[ \bm E \cdot (\bm x \times \bm \Omega)(5+4\tau^2 \Omega^2)-4\tau^2  (\bm \Omega \times \bm E) \cdot (\bm x \times \bm \Omega) \right].
\label{j2pvb0}
\end{eqnarray}
The axial current density has a similar structure:
\begin{eqnarray}
\bm j_{5\ \perp }^{(2)} &=& - \frac{\tau^2 q \chi }{3(1+4 \tau^2 \Omega^2)}\left(\frac{   \partial \bm E}{\partial t}  -2 \tau\bm{\Omega} \times   \frac{\partial \bm E}{\partial t} \right)  - \frac{\mu}{ 2 \pi^2 \hbar^2 } (\bm x \times \bm{\Omega}  ) \times \bm E   \nonumber\\
&&- \frac{q^2 \tau^2}{3 \pi^2 \hbar^2 (1+4 \tau^2 \Omega^2)} \bm E^2  \bm \Omega \left(  1 + \frac{2+32 \tau^2 \Omega^2}{5(1+4\tau^2 \Omega^2)}  \right)
\nonumber\\
&&+ \frac{\tau^2}{5 \pi^2 \hbar^2}  \mu \bm \Omega (\bm \nabla \cdot \bm E) \left( \frac{2}{3} - \frac{7}{30(1+4\tau^2 \Omega^2)}- \frac{1 -4 \tau^2 \Omega^2}{(1+4 \tau^2 \Omega^2)^2}  \right)
\nonumber\\
&&+ \frac{ \tau \mu}{15 \pi^2 \hbar^2(1+4\tau^2 \Omega^2)^2}\bm \Omega \left[ \bm E \cdot (\bm x \times \bm \Omega)(5+4\tau \Omega^2)-4\tau^2 (\bm \Omega \times \bm E) \cdot (\bm x \times \bm \Omega)  \right].
\label{j2pab0}
\end{eqnarray}
We have employed $\bm \nabla  \times \bm E =0$  which follows from  the Maxwell equations, (\ref{Max}),  in the absence of magnetic field.
Let now $\bm E$ and  $\bm \Omega$  be  parallel to each other. The second order vector  current density  is obtained as
\begin{eqnarray}
\bm j_\parallel^{(2)}  &=& -\frac{\tau^2 q \chi }{3(1+4 \tau^2 \Omega^2)}\left(  \frac{\partial \bm E}{\partial t} +4 \tau^2 \bm{\Omega}   \frac{\partial \bm E}{\partial t}\cdot \bm{\Omega}\right) + \frac{ \mu_5 }{ 2 \pi^2 \hbar^2 }(\bm x \times \bm{\Omega})  \times \bm E
\nonumber\\
&&- \frac{  \mu_5 \tau^2}{5 \pi^2 \hbar^2 (1+4\tau^2 \Omega^2)} (\bm \nabla \cdot \bm E) \bm \Omega \left( \frac{5}{2}+\frac{16 \tau^2 \Omega^2}{3} + \frac{1}{1+4 \tau^2 \Omega^2}       \right)
\nonumber\\
&&-\frac{4 \tau^2 \mu_5(3 + 8 \tau^2 \Omega^2) }{15 \pi^2 \hbar^2 (1+4\tau^2 \Omega^2)^2}\Omega^2  (\bm x \cdot \bm E)\bm \Omega   -  \frac{2 \tau \mu_5}{15 \pi^2 \hbar^2}  E  \Omega (\bm x \times \bm \Omega)  .
\label{j2dvb0}
\end{eqnarray}
Obviously, the second order axial  current density has the similar form
\begin{eqnarray}
\bm j_{5\ \parallel }^{(2)} &=&-\frac{\tau^2 q \chi_5 }{3(1+4 \tau^2 \Omega^2)}\left(  \frac{\partial \bm E}{\partial t} +4 \tau^2 \bm{\Omega}   \frac{\partial \bm E}{\partial t}\cdot \bm{\Omega}\right)+ \frac{\mu}{ 2 \pi^2 \hbar^2 } (\bm x \times \bm{\Omega})  \times \bm E \nonumber\\
&&- \frac{\tau^2}{15 \pi^2 \hbar^2 (1+4\tau^2 \Omega^2)^2}E^2 \bm \Omega \left( 3+16 \tau^2 \Omega^2  - 104 \tau^4 \Omega^4 -64\tau^6\Omega^6  \right)   \nonumber\\
&&- \frac{ \mu \tau^2 }{5 \pi^2 \hbar^2 (1+4\tau^2 \Omega^2)} (\bm \nabla \cdot \bm E) \bm \Omega \left(\frac{5}{2} + \frac{16 \tau^2 \Omega^2}{3} + \frac{1}{1+4 \tau^2 \Omega^2}       \right)
\nonumber\\
&&-\frac{4 \tau^2 \mu }{15 \pi^2 \hbar^2 (1+4\tau^2 \Omega^2)^2}\Omega^2 \bm \Omega (3 + 8 \tau^2 \Omega^2) \bm E \cdot \bm x \nonumber\\
&&-  \frac{2  \mu \tau}{15 \pi^2 \hbar^2}E\Omega (\bm x \times \bm \Omega) .
\label{j2dab0}
\end{eqnarray}
Some comments are in order.
When  electric field does not depend explicitly on the direction of local position, all the explicitly $\bm x$ dependent terms  vanish when one considers the total current $\bm j_t\equiv \int d^3x \bm j.$
In a frame moving with the velocity $\bm v =\bm \Omega \times \bm{x},$ the magnetic field  appears as $\bm B^\prime \approx \bm B-\bm v \times \bm E,$  at the leading order in the Lorentz transformations. Although  we have set $\bm B=0,$ as the reminiscent  
of Lorentz transformation we obtain the last terms in the first lines of (\ref{j2pvb0})-(\ref{j2dab0}).  Actually the $\bm x$ dependent terms are  related to the linear velocity $\bm v$ of the rotating frame.
The first two terms of  (\ref{j2pvb0})-(\ref{j2dab0}), which are linear in the time derivatives of the electric field generate current densities 
which are the counterparts of the electric field dependent terms  in  (\ref{jj11}) and (\ref{5511}).  
One can observe that  in the vector current densities   (\ref{j2pvb0}) and (\ref{j2dvb0}) only the terms proportional to  $\partial \bm E/ \partial t$
survive for a vanishing axial chemical potential. In the axial current densities (\ref{j2pab0}) and (\ref{j2dab0}) the same situation occurs when the total chemical potential vanishes.

The current densities generated by the gradients of chemical potentials 
behave mostly like the current densities generated by the electric field. However there are also nonlinear currents which are proportional to
$\bm \nabla \mu \times \bm E,,\ \bm \nabla \mu_5 \times \bm E$ and  $\bm \Omega (\nabla \mu)^2,\ \bm \Omega (\nabla \mu_5)^2.$

\section{The weak fields approximation}
\label{wfa}

For
small relaxation time which corresponds to large collision rate the solutions of  kinetic equation (\ref{f1o}), (\ref{f2o}) simplify considerably. Now,  besides $\bm E$ one can also regard  $\bm \Omega$ and $\bm B$  as perturbations  \cite{KrTr}.   We would like to calculate the current densities and the evolution equations in time of the chiral chemical potentials under these assumptions. We  deal with the
electromagnetic fields, angular velocity and the derivatives of chemical potentials up to second order.  As usual 
the derivatives of  electromagnetic fields are viewed  as second order. One
can easily read the first order distribution function from  (\ref{f1o}) as 
\begin{eqnarray}
(f^{(1)}_\ssW)_\ssl^i &=&  \frac{\tau  D^i_\ssl}{T}	\hat{\bm{p}}\cdot 
\left(q_i\bm E -\bm \nabla {\mu_\ssl}\right), \nonumber 
\end{eqnarray}
where we set $\partial {\mu_\ssl} / \partial t =0,$ which follows from the consistency condition (\ref{con})  as it can be deduced from (\ref{coc1}). Therefore,   $\partial {\mu_\ssl} / \partial t$ is handled  as  second order. 
Both  (\ref{f1o}) and (\ref{f2o}) possess terms which are  second order in $\bm E,\ \bm B,\ \bm \Omega, \bm \nabla \mu_\ssl,$   which can be shown to yield
\begin{eqnarray}
(f^{(2)}_\ssW)^i_\ssl
&=&  \frac{\tau  q_i D^i_\ssl}{T}	
\hat{\bm{p}}\cdot\left[ 
- g \bm{A}_i \times \bm E_{\mu_\ssl}^i  + \bm b^\lambda_i  ( \bm E_{\mu_\ssl}^i \cdot \bm{A}_i) 
-	\bm E_{\mu_\ssl}^i (\bm b^\lambda_i \cdot \bm{A}_i) \right]-\frac{\tau D^i_\ssl}{T}\frac{\partial (\bm \chi^0_{i\ssl} \cdot \bm p)}{\partial t} 	 \nonumber \\
&&
- \frac{\tau D^i_\ssl}{T^2} \left(  q_i(1-2 f^0)  (\bm \chi^0_{i\ssl} \cdot \bm p) ( \hat{\bm p} \cdot \bm E_{\mu_\ssl}^i) +  T  \frac{\partial (\bm \chi^0_{i\ssl}  \cdot \bm p)}{\partial \bm p} \cdot \bm e + T   \hat{\bm p} \cdot  \frac{\partial (\bm \chi^0_{i\ssl} \cdot \bm p)}{\partial \bm x} \right) \nonumber \\
&&- \frac{\tau D^i_\ssl }{T}\frac{\partial \mu_\ssl}{\partial t} + \frac{ q_i\tau D^i_\ssl }{T} (\hat{\bm p} \cdot \bm E_{\mu_\ssl}^i)(\hat{\bm p} \cdot (\bm x \times \bm \Omega)) - \frac{ q_i\tau D^i_\ssl}{T} (\bm E \times \bm b^\lambda_i) \cdot  \bm \nabla \mu_\ssl ,
\end{eqnarray}
where
$\bm \chi^0_{i\ssl}$ is defined by  (\ref{chi0}) after substituting $q$ and $\mu$ with $q_i$ and $\mu_\ssl .$ 

The consistency condition (\ref{con}) for $(f^{(1)}_\ssW)^i_\ssl +(f^{(2)}_\ssW)^i_\ssl$ can be computed as follows
\begin{eqnarray}
&& \frac{ 3 {\mu_\ssl}^2 +\pi^2 T^2 }{6 \pi^2\hbar^3} \frac{\partial \mu_\ssl}{\partial t} 
 +\frac{\mu_\ssl ({\mu_\ssl}^2 +\pi^2 T^2 ) }{3 \pi^2\hbar^3} \bm \Omega^2   - \frac{ \lambda q }{4 \pi^2\hbar^2}  {\bm E_\mu}_\ssl \cdot (q\bm B + 2\mu_\ssl  \bm \Omega) \nonumber\\
&& +\frac{ q \tau }{3 \pi^2\hbar^3} \mu_\ssl {\bm E_\mu}_\ssl \cdot \bm \nabla \mu_\ssl +\frac{q \tau (3 {\mu_\ssl}^2 +\pi^2 T^2 )}{18 \pi^2\hbar^3} \bm \nabla \cdot \bm E^\prime_{\mu_\ssl}  =0 , \label{ccw}
\end{eqnarray}
where $\bm E^\prime_{\mu_\ssl}=\bm E + (\bm\Omega\times \bm x)\times \bm B -\frac{1}{q}\bm \nabla \mu_\ssl .$ 
The chiral current density up to second order  is obtained  as
\begin{eqnarray}
\bm j^{\ssW}_{\ssl}&=&\frac{q \lambda}{4 \pi^2\hbar^2} \mu_\ssl \bm B + \frac{\lambda(3 {\mu_\ssl}^2 +\pi^2 T^2 )}{12 \pi^2\hbar^3} \bm \Omega	+ \frac{ q \lambda \mu_\ssl }{4 \pi^2\hbar^2} (\bm x \times \bm \Omega) \times \bm E \nonumber\\
&&+ \frac{\tau q(3 {\mu_\ssl}^2 +\pi^2 T^2 ) }{18 \pi^2\hbar^3}{\bm E_\mu}_\ssl - \frac{q \tau^2 (3 {\mu_\ssl}^2 +\pi^2 T^2 )}{18 \pi^2\hbar^3} \frac{\partial \bm E }{\partial t}	\nonumber\\
&&- \frac{q^2 \tau^2}{6 \pi^2\hbar^3} \mu_\ssl (\bm B \times{\bm E_\mu}_\ssl) - \frac{q \tau^2 (3 {\mu_\ssl}^2 +\pi^2 T^2 )}{9 \pi^2\hbar^3} (\bm \Omega \times {\bm E_\mu}_\ssl)\nonumber\\
&&+ \frac{q \tau (3 {\mu_\ssl}^2 +\pi^2 T^2 )}{18 \pi^2\hbar^3} (\bm \Omega \times \bm x) \times \bm B + \frac{\tau \mu_\ssl ( {\mu_\ssl}^2 +\pi^2 T^2) }{6 \pi^2\hbar^3}(\bm \Omega \times \bm x) \times \bm \Omega\nonumber\\
&&+\frac{\tau \lambda q}{12 \pi^2 \hbar^2} (\mu_\ssl \bm \nabla \times \bm E -  \bm E \times \bm \nabla \mu_\ssl )  . \label{jla}
\end{eqnarray}
Here the last line comes from the magnetization current (\ref{curl}). The vector and axial  currents can be acquired by inserting  (\ref{jla}) into (\ref{eac}).

\subsection{Currents and the continuity equations}
By summing (\ref{jla}) over $\lambda$ one establishes the vector  current  as
\begin{eqnarray}
\bm j^{\ssW} & = &
\frac{q\mu_5}{2 \pi^2\hbar^2}   \bm B + \frac{\hbar\chi_5}{2}  \bm \Omega - \frac{ q \mu_5 }{2 \pi^2\hbar^2} (\bm \Omega\times \bm x) \times \bm E   \nonumber\\
&&+ \frac{\tau q \chi }{3} \bm E -\frac{\tau \chi}{3 }  \bm \nabla \mu -\frac{\tau \chi_5}{3 }  \bm \nabla \mu_5  - \frac{\tau^2 q \chi }{3} \frac{\partial \bm E}{\partial t} \nonumber\\
&&- \frac{q^2 \tau^2 \mu}{3 \pi^2\hbar^3}  \bm B \times \bm E + \frac{q \tau^2}{6} \bm B \times \bm \nabla \chi    \nonumber\\
&&- \frac{ 2q \tau^2 \chi}{3} \bm \Omega \times \bm E + \frac{ 2 \tau^2 }{3 } \bm \Omega \times \bm \nabla n \nonumber\\
&&+ \frac{ q \tau \chi }{3} (\bm \Omega \times \bm x) \times \bm B + \tau n(\bm \Omega \times \bm  x) \times \bm  \Omega
\nonumber\\
&&+\frac{\tau q }{6 \pi^2 \hbar^2} (\mu_5  \bm \nabla \times \bm E - \bm E \times \bm \nabla \mu_5) 
. \label{wvc}
\end{eqnarray}      
On the other hand the summation of    (\ref{ccw}) over $\lambda $ yields the condition 
\begin{eqnarray}
&& + \chi_5 \frac{\partial \mu_5}{\partial t} + \chi \frac{\partial \mu}{\partial t}   +\frac{ q }{2 \pi^2\hbar^2} \bm B \cdot \bm \nabla \mu_5+ \frac{\hbar}{2} \bm \Omega \cdot  \bm \nabla \chi_5   - \frac{ q \mu_5}{ \pi^2\hbar^2}  \bm E \cdot \bm \Omega  \nonumber\\
&&+ \frac{q \tau \chi}{3}  \bm \nabla \cdot \bm E + \frac{q \tau \chi}{3}  \bm \Omega\cdot\bm B -\frac{\tau \chi}{3}   \bm \nabla^2 \mu 
-\frac{\tau \chi_5}{3} {\bm \nabla}^2 \mu_5  +2 \tau n \bm\Omega^2 \nonumber\\
&&  +\frac{q \tau}{3} \bm E \cdot \bm \nabla \chi - \frac{ \tau }{3}  \bm \nabla \chi \cdot \bm \nabla \mu 
- \frac{ \tau }{3}  \bm \nabla \chi_5 \cdot \bm \nabla \mu_5 = 0
 .\label{ccwv}
\end{eqnarray}
One can check that by virtue of (\ref{ccwv})  the electric   charge is conserved:
$$
\frac{\partial n}{\partial t}+\bm \nabla \cdot \bm j^{\ssW} =0.
$$ 
Recall that we use particle number and current densities, so that  the electric charge  and current densities are  given by $qn$ and $q\bm j^\ssW.$

By plugging  (\ref{jla}) into  (\ref{eac}),   the axial  current can be shown to be
\begin{eqnarray}
\bm j^{\ssW}_5 & = &
\frac{q   \mu}{2 \pi^2\hbar^2}    \bm B+ \frac{\hbar \chi}{2}  \bm \Omega  - \frac{ q  \mu }{2 \pi^2\hbar^2} (\bm \Omega \times \bm x) \times \bm E  \nonumber\\
&&+  \frac {\tau q\chi_5}{3}  \bm E -  \frac{\tau \chi_5 }{3}  \bm \nabla \mu -  \frac{\tau \chi }{3}  \bm \nabla \mu_5  - \frac {\tau q\chi_5}{3}  \frac{\partial \bm E}{\partial t}
\nonumber\\
&&- \frac{ \tau^2 q^2 \mu_5 }{3 \pi^2\hbar^3} \bm B \times \bm E +  \frac{q \tau^2}{6}   \bm B \times \bm \nabla \chi_5    \nonumber\\
 &&- \frac{2 q \tau^2 \chi_5}{3} \bm \Omega \times \bm E + \frac{  2\tau^2 }{3 } \bm \Omega \times \bm \nabla n_5 \nonumber\\
&&+ \frac{  q \tau \chi_5}{3 }     (\bm \Omega \times \bm x) \times \bm B  + \tau n_5 (\bm \Omega \times \bm  x) \times \bm  \Omega
\nonumber\\
&&+\frac{\tau q}{6 \pi^2 \hbar^2} (\mu  \bm \nabla \times \bm E - \bm E \times \bm \nabla \mu) . \label{wac}
\end{eqnarray}  
Equation (\ref{ccw}) in addition to (\ref{ccwv}) leads to the  condition 
\begin{eqnarray}
&& \chi   \frac{\partial \mu_5}{\partial t}+ \chi_5 \frac{\partial \mu}{\partial t} +\frac{ q }{2 \pi^2\hbar^2}  \bm B \cdot \bm \nabla \mu + \frac{\hbar}{2 } \bm \Omega  \cdot \bm \nabla \chi - \frac{ q \mu }{ \pi^2\hbar^2}  \bm E \cdot \bm \Omega -\frac{ q^2 }{2 \pi^2\hbar^2} \bm B \cdot {\bm E}  \nonumber\\
&& +\frac{q \tau \chi_5}{3}  \bm \nabla \cdot \bm E + \frac{q \tau \chi_5}{3}  \bm \Omega\cdot\bm B  -\frac{ \tau \chi_5}{3}  \bm \nabla^2  \mu - \frac{\tau \chi}{3} {\bm \nabla}^2 \mu_5 + 2 \tau  n_5  \bm \Omega^2   \nonumber\\
&& +\frac{q \tau}{3} \bm E \cdot  \bm \nabla \chi_5  - \frac{ \tau }{3}  \bm \nabla \chi \cdot \bm \nabla \mu 
- \frac{ \tau }{3}  \bm \nabla \chi_5 \cdot \bm \nabla \mu_5
= 0, \label{ccwa} 
\end{eqnarray}
which  is acquired by  multiplying (\ref{ccw}) with $\lambda$ and then summing over it, similar to the definition of the chiral current (\ref{eac}).
By making use of the consistency condition (\ref{ccwa}), one can show that the chiral 4-current satisfies the  anomalous continuity equation
$$
\frac{\partial n_5}{\partial t}+\bm \nabla \cdot \bm j_5^{\ssW} =  \frac{  q^2}{2\pi^2 \hbar^2} \bm E \cdot \bm B.
$$ 
The time evolution equations of the total and chiral chemical potentials, $\mu,\mu_5,$ are given by the coupled equations (\ref{wvc}) and (\ref{wac}).

For $\bm \Omega=0$ we obtain the results reported in Ref.\cite{gorbaretal}.

The first two terms in (\ref{wvc}) are the chiral magnetic and vorticity effects. In (\ref{wac}) the first and second terms, respectively,  are the chiral separation and local polarization effects.
In a frame moving with the velocity $\bm v,$  the Lorentz transformed magnetic field can be approximated as $\bm B^\prime \approx \bm B -\bm v \times \bm E.$ Actually 
the first and third terms in (\ref{wvc}) and (\ref{wac}) are exactly in this form, where the linear  velocity is due to rotation: $\bm v=\bm \Omega \times \bm x .$ 

For the linear terms in $\bm E$ we can adopt the discussions given in Section \ref{b0}: First let $\bm E$ be parallel to both $\bm B$ and $\bm \Omega.$ 
Then as before  the vector current density yields the
Ohm's law,
$\bm j_{\mathrm{Ohm}} =(q\tau\chi/3)\bm E,$ and the chiral current density leads to the
electric separation effect, $ \bm j_{\mathrm{sep}}=(q\tau\chi_5/3)\bm E.$ When the magnetic field and angular velocity are both in the $z-$direction, we obtain  the currents (\ref{Jx})-(\ref{Jy}) but now $ a_1, a_2 $ are  given by
$a^\ssV_1= q\tau\chi/3,\ a^\ssV_2= \frac{q^2 \tau^2 \mu}{3 \pi^2\hbar^3} B+ \frac{ 2q \tau^2 \chi}{3} \Omega$  for the vector current density (\ref{wvc}) and by $a^\ssA_1 =q\tau \chi_5 /3,\ a^\ssA_2= \frac{q^2 \tau^2 \mu_5}{3 \pi^2\hbar^3} B+ \frac{ 2q \tau^2 \chi_5}{3} \Omega$ for the axial current density   (\ref{wac}). Hence, the conductivity 
$$
\sigma^{(\Omega, B)}_\ssH =
\frac{  q \chi^2+\tau^2\left(\frac{q \mu}{\pi^2\hbar^3} B+  2\chi\Omega\right) ^2}{\frac{3q\mu}{\pi^2\hbar^3} B+ 6\chi \Omega}
$$ 
is independent of   the relaxation time in the limit $\tau\ll 1: $ 
$$
\sigma^{(\Omega, B)}_\ssH \approx
\frac{   \chi^2}{\frac{3q\mu}{\pi^2\hbar^3} B+ 6\chi \Omega} .
$$ 
This is the Hall conductivity associated to the effective magnetic field $ \bm B_{eff}= \bm B +2(\chi/\mu)\bm \Omega.$
The same discussion can be done for the axial current  (\ref{wac}) by substituting $\chi$ and $\mu$
with $\chi_5$ and $\mu_5.$  The last terms of the third and fourth lines of  (\ref{wvc}) and  (\ref{wac})  
obviously yield Hall-like currents associated to the gradients of the susceptibilities $\bm \nabla \chi, \bm \nabla \chi_5$ and the number densities $\bm \nabla n, \bm \nabla n_5,$ instead of the electric field $\bm E.$ The second and third terms in the second line   of  (\ref{wvc}) and  (\ref{wac})  are   diffusion current densities. 
The first terms of the last lines  of (\ref{wvc}) and  (\ref{wac}) which depend  on both $\bm \Omega$ and $\bm B,$ combined with  the
first terms of the second lines which are linear in $\bm E ,$ can be written in terms of the  electric   field in rotating coordinates, 
$\bm E^\prime =  \bm E + (\bm\Omega\times \bm x)\times \bm B .$ The last terms in the fifth lines of (\ref{wvc}) and (\ref{wac}) are clearly due to the fictitious  centrifugal force. Magnetization current produces  the last two terms of (\ref{wvc}) and (\ref{wac}). 

\section{Conclusions}

In coordinates rotating with a constant angular velocity the kinetic equation of charged particles in the presence of electric and magnetic fields is  studied by means of the relaxation time approach. We solved the Boltzmann transport equation approximately up to  second order in the  electric  field expressed in rotating coordinates. We calculated currents following from the collision terms for hot chiral plasma by taking into account both particles and antiparticles.  We considered first the chiral plasma for vanishing magnetic field. We found that there are several new terms as well as some known ones due to Coriolis and centrifugal forces. We then studied chiral plasma for weak fields in the small relaxation time regime. We showed that angular velocity generate currents similar to magnetic fields. 
Collision terms should respect the particle number, which is guaranteed in terms of the consistency condition (\ref{con}).  
It gives the equations which should be satisfied by the inhomogeneous kinetic potential and electromagnetic fields which evolve according to the Maxwell equations in rotating frames, (\ref{Max}). In fact these conditions are essential to show that the particle 4-currents obey the continuity equations with the chiral anomaly. 

\begin{acknowledgments}
	This work is supported by the Scientific and Technological Research Council of Turkey (T\"{U}B\.{I}TAK) Grant No. 115F108.
		
\end{acknowledgments}

\appendix 

\renewcommand{\theequation}{\thesection.\arabic{equation}}
\setcounter{equation}{0}

\section{Equilibrium distribution function}

Consider the Lorentz scalar equilibrium distribution function (\ref{eqf}) where
 the linear velocity is due to rotation, $\bm v = \bm \Omega \times \bm x,$ and $p_0$ is given by the fully fledged dispersion relation, (\ref{disp}), by expanding it  in Taylor series as
\begin{eqnarray}
f^{i}_{eq \ssl} &=& \frac{1}{e^{[{\cal E}_i^\lambda -\bm p \cdot (\bm \Omega \times \bm x)-\mathrm{sign}  (q_i)\mu_\ssl ]/T} +1}  \nonumber \\
&\approx & f^{0i}_\ssl  -  \frac{\partial f^{0i}_\ssl }{\partial p}\big[  p^2 (\bm{b}_i^\lambda \cdot \bm{\Omega}) + q_ip (\bm{b}_i^\lambda \cdot \bm B  ) + \bm p \cdot (\bm \Omega \times \bm x) \big]\nonumber \\
&&+ \frac{\partial ^2  f^{0i}_\ssl }{\partial p^2} \big[  p^2 (\bm{b}_i^\lambda \cdot \bm{\Omega}) + q_ip (\bm{b}_i^\lambda \cdot \bm B  ) \big] \bm p \cdot (\bm \Omega \times \bm x) .\label{feqT}
\end{eqnarray}
If one would like to use of it in calculating  the chiral number density, one should set $\bm \Omega \times \bm x=0$ in (\ref{S1}) and define
\begin{eqnarray}
n_i^\lambda &= &\int \frac{d^3p}{(2\pi\hbar)^3}[1+  \ \bm{b}_i^\lambda \cdot (q_i\bm{B} + 2p \bm{\Omega} )] f^{i}_{eq \ssl}.
\nonumber
\end{eqnarray}
Its  explicit calculation leads to
\begin{eqnarray}
n_i^\lambda &= &\int \frac{dp}{2\pi^2\hbar^3} \Big\{  p^2f^{0i}_\ssl 
- \frac{q\lambda\hbar}{6} p  \frac{\partial f^{0i}_\ssl }{\partial p} \bm B\cdot (\bm \Omega \times \bm x)
\nonumber \\
&&+  \frac{q\lambda\hbar}{6}\frac{\partial ^2  f^{0i}_\ssl }{\partial p^2} p^2 \bm B \cdot (\bm \Omega \times \bm x) \Big\}  \nonumber \\
&&=\frac{\mu_\ssl }{6 \pi^2\hbar^3} (\mu_\ssl ^2 + \pi^2 T^2)+ \frac{q\mu_\ssl}{4 \pi^2\hbar^2}  \bm B \cdot (\bm \Omega \times \bm x).
\end{eqnarray}
It is the same with the number density (\ref{nll}), which was  obtained from  (\ref{S1}) after setting ${\bm \nu}_{0i}^\lambda =p$ and by taking $f^{0i}_\ssl$ as the equilibrium distribution function. It is worth noting that even if $\bm \Omega =0,$   for particles moving with the velocity $\bm v,$ in the number density  there will be the term   $  \frac{q\mu_\ssl}{4 \pi^2\hbar^2}  \bm B \cdot \bm v,$ due to the equilibrium distribution 
$$
f^{i}_{eq \ssl} = \frac{1}{e^{ [ 1-q_ip (\bm{b}_i^\lambda \cdot \bm B  ) -\bm p \cdot \bm v-\mathrm{sign}  (q_i)\mu_\ssl ]/T} +1} ,
$$
and
$
(\sqrt{\omega})_i^\lambda =1+   q_i\bm{b}_i^\lambda \cdot\bm{B} .
$

We would like to clarify that our formalism takes into account rotation correctly when we employ 
 $f^{0i}_\ssl .$  In \cite{cssyy} for $\bm B=\bm E=0,$  the current is defined as 
$$
\bm J = \sum \limits_i \mathrm{sign}(q_i) \int \frac{d^3p}{(2\pi\hbar)^3}\left( \hat{\bm p}f^{i}_{eq \ssl}-\frac{\hat{\bm p}}{2p}\times \bm \nabla f^{i}_{eq \ssl} \right),
$$
where the latter term is due to magnetization. Now, by keeping the first order terms in (\ref{feqT})  for small $\bm \Omega \times \bm x,$ one obtains
$$
\bm J= \frac{\lambda (3 {\mu_\ssl}^2 + \pi^2 T^2)}{12 \pi^2\hbar^2 }  \bm \Omega .
$$
It is the same with  chiral magnetic effect which we reported  in (\ref{jlf0}), obtained by (\ref{S2}) and $f^{0i}_\ssl .$ 

Let us now show that using energy (\ref{disp}) and (\ref{feqT}) or  $p_0=p$ and $f^{0i}_\ssl $  yield the same 
current  when the system is in equilibrium.  In equilibrium  Lorentz and centrifugal  forces vanish: $\bm e=0.$ 
Using $\cal E$ and (\ref{feqT})  for $\bm e=0$ and  vanishing  centrifugal term, $(\bm \Omega \times \bm x)^2=0,$ (\ref{jil})   yields 
\begin{eqnarray}
\bm j_i^\lambda  &=&   \int \frac{d^3p}{(2\pi\hbar)^3} \{ {\bm \nu}_{0i}^\lambda +
(\bm{b}_i^\lambda \cdot \hat{\bm p}) (q_i\bm{B} + 2p \bm{\Omega} ) \} f^{i}_{eq \ssl} \nonumber \\
&& =\int \frac{ 4 \pi dp}{(2\pi\hbar)^3} \hbar \lambda sign(q_i) \bm \Omega  \big\{ \frac{2 p}{3} f^{0i}_\ssl  - \frac{p^2 }{6} \frac{\partial f^{0i}_\ssl }{\partial p}    \big\}
 \nonumber \\
&&+ \int \frac{ 4 \pi dp}{3 (2\pi\hbar)^3} \hbar \lambda q \bm B \big\{  f^{0i}_\ssl - \frac{p}{2} \frac{\partial f^{0i}_\ssl }{\partial p} 
  \big\} \nonumber \\
&&=\frac{\lambda(3 {\mu_\ssl}^2 +\pi^2 T^2 )}{12 \pi^2\hbar^3} \bm \Omega + \frac{q \lambda}{4 \pi^2\hbar^2} \mu_\ssl \bm B ,
\end{eqnarray}
 for small $\bm \Omega \times \bm x.$  It is equal to the first two terms of (\ref{jla}), which were obtained by $p$ and $ f^{0i}_\ssl .$

\section{Fermi-Dirac integrals}

The Fermi-Dirac integral of integer order is defined in terms of the Fermi-Dirac distribution as
$$
F_k(\mu)\equiv\int_{0}^{\infty}dp p^k f^0 .
$$
For $k\neq 0$ one can easily observe that
\begin{equation}
\label{kkmo}
\frac{dF_k(\mu)}{d\mu}=k F_{k-1}(\mu).
\end{equation}
Integrals are performed in terms of the gamma functions and polylogarithms (see \cite{fuk} and the references therein) as
\begin{equation}
\label{reli}
F_k(\mu)=-T^{k+1} k!\ \Li_{k+1}(-e^{\mu/T}), \ \ k=-1,0,1\cdots 
\end{equation}
The polylogarithms $\Li_s$ can be defined as series expansion  $\Li_s(x)=\sum_{n=1}(x^n/n^s),$ where  
$$
\Li_0(x)=\frac{x}{1-x},\ \ \ \Li_1 =-\ln (1-x).
$$
From (\ref{kkmo}) and (\ref{reli})  we acquire the integrals which we need:
\begin{eqnarray*}
	\int_{0}^{\infty}dp p^k D_\ssl   &=&-T^{k+1} k!\ \Li_{k}(-e^{\mu/T}), \ \ k=0,1,\cdots  \\	
	\int_{0}^{\infty}dp p^k D_\ssl (1-2f^0)  &=&-T^{k+1} k!\ \Li_{k-1}(-e^{\mu/T}), \ \ k=1,2,\cdots , 
\end{eqnarray*}
where
$$
\frac{df^0}{d p}=-\frac{d f^0}{d \mu}=-\frac{1}{T} D_\ssl
$$
Some relations which we use in our calculations  are:
\begin{eqnarray*}
	&\frac{1}{1+e^x} +\frac{1}{1+e^{-x}}=1 &\\
	&\ln (1+e^{x}) - \ln (1+e^{-x}) =x \\
	&\Li_2 (-e^{x})+\Li_2 (-e^{-x}) =-\frac{\pi^2}{6}-\frac{x^2}{2}&\\
	&\Li_3 (-e^{x})-\Li_3 (-e^{-x}) =-\frac{x^3}{6} -\frac{\pi^2x}{6}&\\
	&\Li_4 (-e^{x})+\Li_4 (-e^{-x}) = -\frac{7\pi^4}{360}-\frac{x^4}{24} -\frac{\pi^2x^2}{12}.&
\end{eqnarray*}

\section{The centrifugal terms for $\bm B=0$ }

The  chiral current densities given by the centrifugal force dependent terms of $f^2$ are calculated as
\begin{eqnarray}
\bm j_\ssl (f^2) &=& \frac{\lambda \tau \bm \Omega}{\pi^2\hbar^3} \Bigg\{  \frac{1-8 \tau^2 \Omega^2}{30(1+4 \tau^2 \Omega^2)} \left[\mu_\ssl K (\bm E_{\mu_\ssl}, \Omega^2) + \frac{(\pi^2 T^2 + 3 \mu_\ssl ^2)}{3} K ( \Omega ^2, \Omega ^2)\right]\nonumber\\
&+&   \frac{3+32 \tau^2 \Omega^2}{30(1+4 \tau^2 \Omega^2)} \left[\mu_\ssl L (\bm E, \Omega^2) + \frac{(\pi^2 T^2 + 3 \mu_\ssl ^2)}{3} L ( \Omega ^2, \Omega ^2)\right]\nonumber\\
&-&  \frac{\tau}{5(1+4 \tau^2 \Omega^2)} \left[\mu_\ssl M(\bm E_{\mu_\ssl}, \Omega^2) + \frac{(\pi^2 T^2 + 3 \mu_\ssl ^2)}{3} M ( \Omega ^2, \Omega ^2)\right]\nonumber\\
&-&  \frac{\tau}{3(1+4 \tau^2 \Omega^2)} \left[\mu_\ssl N(\bm E, \Omega^2) + \frac{(\pi^2 T^2 + 3 \mu_\ssl ^2)}{3} N ( \Omega ^2, \Omega ^2)\right] \Bigg\} \nonumber\\
&+&  \frac{\hbar \lambda \tau^2 }{30 \pi^2 \hbar^3 (1+4 \tau^2 \Omega^2)} \mu_\ssl \bm H(\bm E_{\mu_\ssl}, \Omega^2) - \frac{2 \hbar \lambda \tau^2}{\pi^2 \hbar^3(1+4 \tau^2 \Omega^2)}\mu_\ssl  \bm T(\bm E_{\mu_\ssl}, \Omega^2) \nonumber\\
&-&  \frac{7 \hbar \tau^2}{30 \pi^2 \hbar^3} \mu_\ssl [(\bm \Omega \times \bm x) \times \bm \Omega] (\bm E_{\mu_\ssl}  \cdot \bm \Omega) + \frac{4 \hbar \lambda \tau^5  \Omega^4}{30 \pi^2 \hbar^3(1+4 \tau^2 \Omega^2)} (\pi^2 T^2 + 3 \mu_\ssl ^2) \bm \Omega \nonumber\\
&-& \frac{\hbar \tau}{6 \pi^2 \hbar^3} \mu_\ssl ({\mu_\ssl}^2 + \pi^2 T^2) \frac{\partial  \tilde{\bm \chi}^0_\ssl [\Omega^2] }{\partial t} +  \frac{\hbar \lambda \tau}{12 \pi^2 \hbar^3} \mu_\ssl [(\bm \Omega \times \bm x) \times \bm \Omega] \times \bm \nabla \mu_\ssl \nonumber\\
&& - \frac{\tau \lambda \hbar }{3 \hbar^3 \pi^4 (1+4 \tau^2 \Omega^2)}\mu_\ssl   \{ [(\bm \Omega \times \bm x) \times \bm \Omega] \times \bm \nabla \mu_\ssl - 2 \tau \Omega^2 (\bm \Omega \times \bm x) \times  \bm \nabla \mu_\ssl  \}.
\nonumber
\end{eqnarray}
The last line is the contribution of the magnetization current (\ref{curl}). We introduced some functionals  which depend on  $$\tilde{\bm \chi}^0 [\Omega^2]= \frac{\tau (\bm \Omega \times \bm x) \times \bm \Omega - 2 \tau^2  \Omega^2 (\bm \Omega \times \bm x)}{1+4 \tau^2 \Omega^2} $$
and   $ \tilde{\bm \chi}^0 [\bm E_{\mu_\ssl}]$ given in (\ref{cto}), as
\begin{eqnarray}
K(\bm E_{\mu_\ssl}, \Omega^2) &=&   \tilde{\bm \chi}^0 [\bm E_{\mu_\ssl}] \cdot  [(\bm \Omega \times \bm x) \times \bm \Omega]  +  \tilde{\bm \chi}^0 [\Omega^2] \cdot \bm E_{\mu_\ssl}, \nonumber \\
K(\Omega^2, \Omega^2) &= &{\tilde{\bm \chi}^0 [\Omega^2]}  \cdot \bm  [(\bm \Omega \times \bm x) \times \bm \Omega]  , \nonumber \\
L(\bm E, \Omega^2)  &= &{\tilde{\bm \chi}^0[\bm E_{\mu_\ssl}]}  \cdot  [(\bm \Omega \times \bm x) \times \bm \Omega]  + {\tilde{\bm \chi}^0 [\Omega^2]}  \cdot \bm E, \nonumber \\
L( \Omega^2, \Omega^2)&=&{\tilde{\bm \chi}^0[\Omega^2]}\cdot [(\bm \Omega \times \bm x) \times \bm \Omega]  , \nonumber \\
M(\bm E_{\mu_\ssl}, \Omega^2) &=& \bm E_{\mu_\ssl} \cdot (\bm \Omega \times   \tilde{\bm \chi}^0 [\Omega^2]   )+ [(\bm \Omega \times \bm x) \times \bm \Omega] \cdot  (\bm \Omega \times \tilde{\bm \chi}^0 [\bm E_{\mu_\ssl}] ) , \nonumber \\
M(\Omega^2, \Omega^2) &= & [(\bm \Omega \times \bm x) \times \bm \Omega] \cdot (\bm \Omega \times \tilde{\bm \chi}^0[\Omega^2]  ) ,
\nonumber \\
N(\bm E, \Omega^2) &=& \bm E \cdot (\bm \Omega \times \bm \tilde{\bm \chi}^0[\Omega^2]    )+ [(\bm \Omega \times \bm x) \times \bm \Omega] \cdot  (\bm \Omega \times \bm \tilde{\bm \chi}^0 [\bm E_{\mu_\ssl}] ) , \nonumber \\
N(\Omega^2, \Omega^2) &=&  [(\bm \Omega \times \bm x) \times \bm \Omega] \cdot (\bm \Omega \times \tilde{\bm \chi}^0[\Omega^2]  ) , \nonumber \\
H(\bm E_{\mu_\ssl}, \Omega^2) &= &\tilde{\bm \chi}^0[\Omega^2] \bm \{ (\bm E_{\mu_\ssl}  \cdot \bm \Omega )(6+12 \tau^2 \Omega^2)+ (\bm E  \cdot \bm \Omega )(1-8 \tau^2 \Omega^2) \bm \} , \nonumber \\
T(\bm E_{\mu_\ssl}, \Omega^2) &= &\bm \Omega \times \tilde{\bm \chi}^0[\Omega^2]  \left(\frac{\bm E_{\mu_\ssl}  \cdot \bm \Omega}{10} + \frac{\bm E  \cdot \bm \Omega}{6}
\right)  . \nonumber
\end{eqnarray}

\newcommand{\PRL}{Phys. Rev. Lett. }
\newcommand{\PRB}{Phys. Rev. B }
\newcommand{\PRD}{Phys. Rev. D }

\end{document}